\documentclass[cits]{PoS}

\usepackage{verbatim}
\usepackage{amsthm}
\usepackage{amsmath}
\usepackage{enumitem}

\newcommand{\ms}{\overline{MS}}

\newcommand{\harxnew}[1]{\href{http://arxiv.org/abs/#1}{\tt #1}}

\title{Locally-Smeared Operator Product Expansions}

\ShortTitle{Smeared Operator Product Expansions}

\author{\speaker{Christopher Monahan}\thanks{Current address: Department of 
Physics and Astronomy, The University of Utah, UT 84112, U.S.A.}\\
Department of Physics, The College of William \& Mary, Williamsburg, 
VA 23185, U.S.A.\\
E-mail: \email{chris.monahan@utah.edu}}

\author{Kostas Orginos\\
Department of Physics, The College of William \& Mary, Williamsburg, 
VA 23185, U.S.A.\\
Thomas Jefferson National Accelerator Facility, Newport News, VA 23606, U.S.A.\\
E-mail: \email{kostas@jlab.org}}

\abstract{
We propose a ``locally-smeared Operator Product Expansion'' (sOPE) to 
decompose non-local operators in terms of a basis of 
locally-smeared operators. The sOPE formally connects nonperturbative matrix 
elements 
of smeared degrees of freedom, determined numerically using the gradient flow, 
to non-local operators in the continuum. The nonperturbative matrix 
elements do not suffer 
from power-divergent mixing on the lattice, provided the smearing scale is kept 
fixed in the continuum limit. The presence of this smearing scale
prevents a simple connection to the standard operator product expansion and 
therefore 
requires the construction of a two-scale formalism. We demonstrate 
the feasibility of our approach using the example of real 
scalar field theory.
}

\FullConference{The 32nd International Symposium on Lattice Field Theory,\\
		23-28 June, 2014\\
		Columbia University New York, NY}
		 
\begin{document}

\section{Introduction}

Deep inelastic scattering (DIS) has been one of the primary experimental and 
theoretical tools used to study the strong force and establish quantum 
chromodynamics (QCD) as the theory of quarks and hadrons. The strong 
dynamics of DIS processes are captured by the hadronic tensor, which 
can be factored into infrared-safe perturbative coefficients and parton 
distribution functions (PDFs). The PDFs, which are the leading contribution to 
the nucleon structure functions in the twist expansion, characterise low-energy 
physics and must be determined nonperturbatively.

PDFs cannot be directly calculated on a Euclidean lattice, because they are 
defined in terms of light-cone matrix elements. So PDFs--which depend on the 
target nucleon, but are independent of the scattering process--are usually 
determined using global analyses 
\cite{Jimenez-Delgado:2014xza}. 
A direct nonperturbative method to compute PDFs would be highly desirable, 
both as an important test of lattice QCD and as a means to
constrain global fits of PDFs in regions that are experimentally 
inaccessible. 

Traditionally, lattice calculations have focussed on the Mellin 
moments of PDFs, determined via matrix elements of local 
twist-2 operators, where twist is the dimension minus the spin of the 
operator, that \emph{can} be directly computed in Euclidean space. 
However, the reduced symmetry of the cubic group induces radiative mixing 
between operators of different spin. Moreover, operators of different mass 
dimension suffer 
from power-divergent mixing, which means that the continuum limit cannot be 
taken. Moments up to the fourth moment can be extracted using 
carefully chosen external momenta, but beyond this power-divergent mixing is 
inevitable.

A method was recently proposed to directly compute PDFs on the lattice 
using 
a large-momentum effective theory \cite{Ma:2014jla}. 
The first lattice calculations were presented in 
\cite{Lin:2014zya}, but there are some unsolved challenges 
\cite{Monahan:inprep}: 
the renormalisation of the nonperturbative matrix 
elements has yet to be fully understood and the practical difficulty of 
the resolution of sufficiently large momenta on the lattices has not been 
addressed. 

Here we propose a new formalism--the ``smeared Operator Product 
Expansion'' (sOPE)-- that removes mixing in the continuum limit and, in 
principle, enables the determination of higher moments of PDFs on the lattice. 
This formalism applies to any lattice calculation that suffers from 
power-divergent mixing. We expand non-local continuum operators in a basis of 
smeared operators. Matrix elements of these 
smeared operators can be determined directly on the lattice and these matrix 
elements require no further renormalisation, up to wavefunction 
renormalisation, 
provided the physical smearing scale 
is kept fixed as the continuum limit is taken.

Smearing has been widely applied in lattice computations to reduce ultraviolet 
fluctuations, partially restore rotational symmetry and systematically 
improve the precision of lattice calculations. In the sOPE, we implement 
smearing 
via the gradient flow, a classical evolution of the original degrees 
of freedom towards the stationary points of the action in a new dimension, the 
flow time 
\cite{Narayanan:2006rf,Luscher:2010iy,Luscher:2013cpa}. The 
gradient flow corresponds to a continuous stout-smearing procedure 
\cite{Morningstar:2003gk} and generally enables the use of smearing lengths 
of only one or two lattice spacings, which is much smaller than typical 
hadronic 
length scales and does not distort the low energy 
physics \cite{Davoudi:2012ya}. Moreover, the gradient flow is computationally 
very cheap to implement.

We demonstrate the feasibility of our approach by studying the sOPE applied to 
scalar field theory. We introduce the gradient flow for scalar fields, the 
simplicity of which facilitates instructive comparison between 
the sOPE and the local OPE. We start with a brief discussion of Wilson's 
formulation of the OPE applied to scalar field theory and 
then outline the sOPE. In Section \ref{sec:phi4} we calculate smeared Wilson 
coefficients and derive 
renormalisation group equations in the small flow time limit. We conclude by 
discussing the application 
of the sOPE to realistic lattice computations of twist-2 matrix 
elements.

\section{The operator product expansion}

The OPE for a non-local operator is well-known, so here we simply introduce the 
notation 
necessary for the following discussions. We write the OPE for a non-local 
operator, ${\cal Q}(x)$, as
\begin{equation}
{\cal Q}(x)  \stackrel{x\rightarrow 0}{\sim} \; \sum_k 
c_k(x,\mu) {\cal O}_R^{(k)}(0,\mu).
\end{equation}
The perturbative Wilson coefficients, $c_k(x,\mu)$, are complex functions that 
capture the short-distance physics associated with the renormalised 
local operator ${\cal O}_R^{(k)}(0,\mu)$, which is a polynomial in the scalar 
field and its derivatives. The renormalisation scale is $\mu$ and the 
free-field mass dimension of the local operator governs the 
leading spacetime dependence of the Wilson coefficients. This equation is 
understood in the weak sense of holding between matrix elements.

The most straightforward example is the time-ordered two-point function, ${\cal 
T}\{ \phi(x)\phi(0)\}$, with spacetime separation $x$. For free scalar field 
theory, the OPE is a Laurent expansion around zero spacetime separation,
\begin{equation}
{\cal T}\{ \phi(x)\phi(0)\} = \frac{1}{4\pi^2x^2}\mathbb{I}+ 
\phi^2(0)  + {\cal O}(x).
\end{equation}

Incorporating interactions, the OPE becomes
\begin{equation}\label{eq:opeint}
{\cal T}\{ 
\phi(x) \phi(0)\} =  \frac{c_{\mathbb{I}}(\mu x,mx)}{4\pi^2x^2}\mathbb{I}+ 
c_{\phi^2}(\mu x,mx)\left[\phi^2(0,\mu) \right]_R  + 
{\cal O}(x),
\end{equation}
where we denote renormalised operators by $[\ldots]_R$ and we have factored out 
the leading spacetime dependence from the Wilson 
coefficients. Radiative corrections generate sub-leading dependence 
on the spacetime separation and, written in this form, the Wilson coefficients 
are dimensionless functions of the spacetime separation $x$, the (renormalised) 
mass $m$, and the renormalisation scale, $\mu$. The ${\cal O}(x)$ 
represents terms of order $x$, up to logarithmic corrections.

\subsection{Our proposal: the \emph{smeared} OPE}

We propose a new expansion in terms of smeared operators, the sOPE:
\begin{equation}
{\cal Q}(x)  \stackrel{x\rightarrow 0}{\sim} \; \sum_k 
d_k(\tau,x,\mu) {\cal S}_R^{(k)}(\tau,0,\mu).
\end{equation}
The smeared Wilson coefficients $d_k(\tau,x,\mu)$ are now functions of three 
scales: the smearing scale, 
$\tau$; the spacetime separation, $x$; and the 
renormalisation scale, $\mu$. The leading spacetime 
dependence of the smeared Wilson coefficients is dictated by the mass 
dimension of the corresponding smeared operator, ${\cal S}_R(\tau,0)$; hence 
the leading spacetime dependence is unchanged. 

Returning to the time-ordered two-point 
function, the sOPE is
\begin{equation}
{\cal T}\{ \phi(x)\phi(0)\} = 
\frac{d_{\mathbb{I}}(\mu x,\mu^2\tau,mx)}{4\pi^2x^2}\mathbb{I}+ 
d_{\rho^2}(\mu x,\mu^2\tau,mx)\rho^2(\tau,0,\mu) 
 + 
{\cal O}(x),
\end{equation}
where we denote smeared fields by $\rho(\tau,x)$. The smeared 
fields are often referred to as ``bulk'' fields and 
the unsmeared fields as ``boundary'' fields and we note that 
the flow time has mass dimension $[\tau] = [m]^{-2}$. The sOPE is only valid 
for small flow time values, just as the OPE only converges for small spacetime 
separations.

\section{\label{sec:phi4}Scalar field theory}

Scalar field theory is a particularly 
straightforward arena in which to apply the sOPE, because we can 
solve the flow time equations exactly and there are no gauge-fixing 
\cite{Luscher:2010iy} or fermion renormalisation 
complications \cite{Luscher:2013cpa}. Moreover, we can view the sOPE for scalar 
fields as a resummation of the local OPE and, although it is not 
necessary for our work, it is interesting to note that 
the OPE converges for Euclidean $\phi^4$-theory in four dimensions 
\cite{Hollands:2011gf}.

We work with $\phi^4$-theory in four-dimensional Euclidean spacetime, defined 
by the action
\begin{equation}
S_\phi[\phi] = \frac{1}{2}\int \mathrm{d}^4x\, \left[(\partial_\nu  
\phi)^2 + m^2\phi^2 + \frac{\lambda}{12}\phi^4\right],
\end{equation}
and introduce the scalar 
gradient flow via the flow evolution equation
\begin{equation}
\frac{\partial \rho(\tau,x)}{\partial \tau}= \partial^2 \rho(\tau,x).
\end{equation}
Here $\partial^2$ is the Euclidean, four-dimensional Laplacian operator.

We impose the Dirichlet 
boundary condition 
$\rho(0,x) = \phi(x)$, so that the full solution is
\begin{equation}
\rho(\tau,x) = \int\mathrm{d}^4y\,\int \frac{\mathrm{d}^4p}{(2\pi)^4} 
\,e^{ip\cdot(x-y)} e^{-\tau p^2} \phi(y) = \frac{1}{16\pi^2\tau^2}
\int\mathrm{d}^4y\,e^{-(x-y)^2/4\tau}\phi(y).
\label{eq:smearedfield}
\end{equation}
This demonstrates explicitly the ``smearing'' effect of the gradient flow: the 
flow time exponentially damps ultraviolet fluctuations.  We can
parameterise the smearing radius by the root-mean-square smearing length, 
$s_{\mathrm{rms}}$, which is given by $s_{\mathrm{rms}}^2 = 8\tau$.

Formally we can write the solution to the flow time equation as $
\rho(\tau,x) = e^{\tau\partial^2}\phi(x)$,
which we can expand for sufficiently small flow times as
\begin{equation}\label{eq:smalltrho}
\rho_R(\tau,x,\mu) = \phi_R(x,\mu)+\tau \left[\partial^2 \phi (x,\mu)\right]_R 
+ \tau^2\left[(\partial^2)^2\phi (x,\mu)\right]_R+{\cal O}(\tau^4).
\end{equation}
Therefore, the (renormalised) smeared fields that 
appear in the smeared operators of the sOPE can be represented by an 
infinite tower of Laplacian operators acting on an unsmeared field. 
Clearly, in this case, the sOPE corresponds to nothing more than a 
reorganisation of the local OPE. 

At this stage it is worth commenting on the small flow-time expansion, which 
has proved an important tool for analysing the gradient flow in several 
different contexts, both perturbatively and nonperturbatively 
\cite{Shindler:2013bia,Luscher:2014kea}. These
studies incorporate a small flow-time expansion of smeared fields in terms of 
local  fields. We can view such an expansion 
as a local OPE in the flow time and thereby relate 
smeared field correlation functions to the 
corresponding renormalised 
correlation functions in the original theory, which would otherwise 
be difficult to compute.

In this work, we take a related approach with an interpretation that is 
tailored to the study of power-divergent mixing in lattice calculations. We do 
not expand the bulk fields in terms of boundary fields, but rather take 
as the fundamental objects of study the (matrix elements of)
fields at positive flow time. We formally 
expand matrix elements of non-local operators at vanishing flow time in terms 
of a basis of ``smeared'' operators at positive, but small, flow time.

\subsection{Computing smeared Wilson coefficients}

\begin{figure}
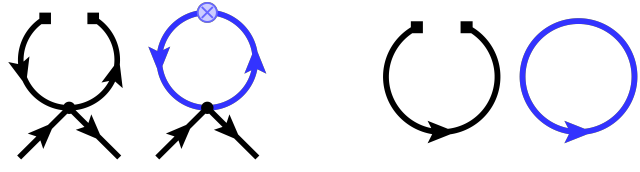

\centering
\begin{minipage}{0.25\textwidth}
\includegraphics[width=0.9\textwidth,keepaspectratio=true]{dphi1loop}
\end{minipage}%
\hspace*{1cm}
\begin{minipage}{0.25\textwidth}
\includegraphics[width=0.9\textwidth,keepaspectratio=true]{d1tree}
\end{minipage}%
\caption{\label{fig:sope2pt} 
Diagrams representing the contributions to the 
smeared Wilson coefficients: $d_{\rho^2}$ to one loop (left-hand 
diagrams) and $d_{\mathbb{I}}$ at tree level (right-hand diagrams). Black and 
blue 
lines are propagators at vanishing and non-vanishing flow times, respectively. 
The black squares are unsmeared fields $\phi(0)$, black dots are 
interaction vertices at vanishing flow time and the blue blob represents 
the smeared operator $\rho^2(\tau,0)$.}
\end{figure}

We show Feynman diagrams that contribute 
to the Wilson coefficient $d_{\rho^2}(\mu x,\mu^2\tau,mx)$ at one loop and 
$d_{\mathbb{I}}(\mu x,\mu^2\tau,mx)$ at tree level in Figure \ref{fig:sope2pt}.
The smeared Wilson coefficient for the leading connected and 
disconnected operators are
\begin{align}\label{eq:drhores}
d_{\rho^2} = {} & 1-\frac{\lambda}{2}\left[\gamma_{\;\mathrm{E}} -1 + 
\log\left(\frac{x^2}{8\tau}\right)\right] +{\cal O}(\lambda^2) \\
d_{\mathbb{I}} = {} & 1 -
\frac{x^2}{8\tau}  +\frac{m^2x^2}{4}\left[\gamma_{\;\mathrm{E}} -1 + 
\log\left(\frac{x^2}{8\tau}\right)\right]+{\cal O}(\lambda)
\end{align}
Here  $\lambda =\lambda_0/(4\pi)^2$, and $\gamma_{\;\mathrm{E}}\simeq 0.577216$ 
is the 
Euler-Mascheroni constant. The corresponding local Wilson coefficients in the 
$\ms$ scheme are
\begin{align}
\!\!\overline{c}_{\mathbb{I}} =  {} & 
1 +\frac{  
m^2x^2}{4} \left[1 +  2\gamma_{\;\mathrm{E}} + 
\log\left(\frac{\mu^2x^2}{16}\right)\right]+{\cal 
O}(\lambda),\\
\!\!\overline{c}_{\phi^2} = {} &  1+\frac{\lambda}{2} \left[1 
+2\gamma_{\;\mathrm{E}}+ 
\log\left(\frac{\mu^2x^2}{16}\right)\right]+{\cal 
O}(\lambda^2). \label{eq:cphires}
\end{align}
Although we have expressed the smeared and local Wilson coefficients in
different renormalisation schemes, so that we cannot directly compare the 
finite contributions, we note four features: 
\vspace*{-1.5\baselineskip}\\
\begin{enumerate}[leftmargin=*]
  \setlength{\itemsep}{1pt}
  \setlength{\parskip}{0pt}
  \setlength{\parsep}{-5pt}
\item The logarithmic dependence on the spacetime separation is the same.
\item The flow time serves as the renormalisation scale for 
the leading order contributions.
\item Wilson coefficients must be independent of the 
external states. We ensure this 
by choosing $\tau< x^2$, or equivalently by taking the flow time 
sufficiently small that terms of ${\cal O}(\tau)$ can be neglected. This
is easily seen 
by considering the derivative with respect to the external momenta of the 
integrand for $d_{\rho^2}$ \cite{Monahan:inprep2}. Choosing the smearing radius
smaller than the spacetime extent of the non-local operator ensures that 
the sOPE remains an (exponentially-)local expansion. In other words, if the 
gradient flow probes length scales on the order of the non-local operator, then
the sOPE becomes a poor expansion for the original operator. This physical 
intuition underlies both the small flow-time expansion 
\cite{Shindler:2013bia,Luscher:2014kea} and the sOPE.
\item The flow time cannot regularise the Wilson coefficients beyond the 
leading order contributions, because the flow evolution is classical and 
interactions necessarily appear at vanishing flow time 
(a higher-order calculation is given in \cite{Monahan:inprep2}). 
However, the 
renormalisation scale dependence of the smeared operators is absorbed by the 
renormalisation parameters of the original theory.
\end{enumerate}

\section{Renormalisation group equations}

The standard renormalisation group (RG) operator is
\begin{equation}
\mu\frac{\mathrm{d}}{\mathrm{d} \mu} = \mu\frac{\partial}{\partial 
\mu}\bigg|_{\lambda,m} + \beta 
\frac{\partial}{\partial \lambda}\bigg|_{\mu,m} - \gamma_m 
m^2\frac{\partial}{\partial m^2}\bigg|_{\mu,\lambda},
\end{equation}
where
\begin{equation}
 \beta = \mu \frac{\mathrm{d}\lambda}{\mathrm{d}\mu}, \qquad
\gamma_m = -\frac{\mu}{2}\frac{\mathrm{d}\log(m^2)}{\mathrm{d}\mu}, 
\;\; \mathrm{and}\;\;
\gamma =  \frac{\mu}{2} \frac{\mathrm{d}\log(Z_\phi)}{\mathrm{d}\mu}.
\end{equation}
Here $Z_\phi$ is the wavefunction renormalisation. In general the beta 
function, $\beta$, and anomalous dimensions, $\gamma$ and $\gamma_m$, are 
functions of the renormalised mass, $m$, the 
renormalisation scale, $\mu$, and the coupling constant, $\lambda$, and are 
known to five loops for an $O(N)$-symmetric theory 
\cite{Kleinert:2001ax}.

For the sOPE, we must account for the extra flow time scale. If we choose 
$\tau = \kappa^2 x^2$, with $\kappa$ a real number that ensures
$s_{\mathrm{rms}}<x$, then the smeared RG operator 
is $\mu\mathrm{d}/\mathrm{d} 
\mu- 2\kappa \mathrm{d}/\mathrm{d} \kappa$.
For small flow times, the smeared Wilson 
coefficient then obeys the smeared RG equation:
\begin{equation}\label{eq:srge}
\left[\mu\frac{\mathrm{d}}{\mathrm{d} \mu}-2\gamma\right]
d_{\rho^2} = 2\left[\kappa\frac{\mathrm{d}}{\mathrm{d} 
\kappa}+\zeta_{\rho^2}\right]
d_{\rho^2},
\end{equation}
where $\zeta_{\rho^2}$ is an anomalous dimension that 
parameterises the flow time dependence of the smeared operator $\rho^2(\tau,0) 
$ \cite{Luscher:2010iy,Luscher:2014kea}. An analogous equation 
holds for the renormalised matrix elements of $\rho^2(\tau,0) 
$, which can be studied nonperturbatively.

\paragraph{Nonperturbative calculations:} We have so far considered the sOPE 
for 
$\lambda \phi^4$ in four spacetime dimensions, which is not asymptotically 
free. Therefore we briefly consider how the sOPE connects 
nonperturbative 
calculations to perturbative results for DIS calculations in QCD. The sOPE 
incorporates two scales, so a ``line of constant physics'' is a path for which 
the scales are tied together, for example, by fixing their product, 
$\mu^2\tau=\kappa^2$. 

Our aim is to connect hadronic matrix elements with smeared Wilson 
coefficients, which are determined in perturbation theory, at a suitably high 
scale using a finite-size scaling analysis 
\cite{Luscher:1991oxn}. We calculate 
matrix elements of smeared operators at some low energy scale, 
determined by the inverse box size, and take the continuum limit at fixed 
physical flow time. The smearing radius should be less than any 
hadronic scales present in the nonperturbative determination of the twist 
operators.

The scaling of the nonperturbative matrix elements is now entirely determined 
by the renormalisation scale dependence and we can 
apply a standard finite volume step-scaling procedure (see, for example, the 
step-scaling procedure for smeared operators in \cite{Monahan:2013lwa}) to some 
high scale. We choose the end point so that the new flow time is small relative 
to the experimentally determined 
inverse momentum transfer of the DIS process. At this new scale, 
the new 
renormalisation scale is sufficiently high that the nonperturbative matrix 
elements 
can be reliably combined with smeared Wilson coefficients.

\section{Summary}

We have proposed a new formalism, which we call the smeared operator product 
expansion (sOPE), to extract lattice matrix elements without power divergent 
mixing. The most obvious application is to deep inelastic scattering, but the 
sOPE can be applied to nonperturbative calculations that suffer 
from power-divergent mixing, such as the computations of $K\rightarrow \pi\pi$ 
decays and $B$-meson mixing \cite{Dawson:1997ic}.  

We expand non-local operators in a basis of smeared operators, generated via 
the gradient flow. The continuum limit of these matrix elements is free from
power-divergent mixing, provided the smearing scale, or flow time, is kept 
fixed in the continuum limit. The matrix elements are then functions of 
both the renormalisation scale and the smearing length. The sOPE 
systematically relates these matrix elements to smeared Wilson coefficients, 
which are calculated perturbatively, and thus provides a complete framework in 
which to extract phenomenologically-relevant physics.

\acknowledgments
We would like to thank Martin L\"uscher and Andrea Shindler for 
enlightening and helpful discussions during 
the course of this work. This project was supported by the
U.S. Department of Energy, Grant Number DE-FG02-04ER41302. K.O.~was also 
supported by the U.S. Department of Energy through Grant Number 
DE-AC05-06OR23177, under which JSA operates the Thomas Jefferson National 
Accelerator Facility. \vspace*{-0.5\baselineskip}


\end{document}